\documentclass[usenatbib]{mn2e}
\pdfoutput=1

\usepackage{natbib}
\usepackage{amsmath}
\usepackage{graphicx}
\usepackage{rotating}
\usepackage{subfig}
\usepackage{amssymb}
\usepackage{multirow}
\usepackage{tabularx}
\usepackage{tabulary}
\usepackage{times}
\usepackage{hyperref}

\newcommand{\kms}{\textrm{ km s}^{-1}}
\newcommand{\dd}{\textrm{d}}

\begin{document}

\title[DG Tau Outflows -- II. Bipolar Outflow Asymmetry]{Multi-epoch Sub-arcsecond [\uppercase{F}\lowercase{e} \uppercase{II}] Spectroimaging of the DG Tau Outflows with NIFS. II. On the Nature of the Bipolar Outflow Asymmetry}

\author[M.~C.~White et al.]
	{M.~C.~White,$^1$
	 G.~V.~Bicknell,$^1$
	 P.~J.~McGregor,$^1$
	 R.~Salmeron$^1$\\
	 $^1$ Research School of Astronomy and Astrophysics, Australian National University, Canberra, ACT, 2611, Australia}
\date{Accepted 2014 April 21.  Received 2014 April 17; in original form 2013 June 26}
\maketitle

\begin{abstract}

The origin of bipolar outflow asymmetry in young stellar objects (YSOs) remains poorly understood. It may be due to an intrinsically asymmetric outflow launch mechanism, or it may be caused by the effects of the ambient medium surrounding the YSO. Answering this question is an important step in understanding outflow launching. We have investigated the bipolar outflows driven by the T Tauri star DG Tauri on scales of hundreds of AU, using the Near-infrared Integral Field Spectrograph (NIFS) on Gemini North. The approaching outflow consists of a well-collimated jet, nested within a lower-velocity disc wind. The receding outflow is composed of a single-component bubble-like structure. We analyse the kinematics of the receding outflow using kinetic models, and determine that it is a quasi-stationary bubble with an expanding internal velocity field. We propose that this bubble forms because the receding counterjet from DG Tau is obstructed by a clumpy ambient medium above the circumstellar disc surface, based on similarities between this structure and those found in the modeling of active galactic nuclei outflows. We find evidence of interaction between the obscured counterjet and clumpy ambient material, which we attribute to the large molecular envelope around the DG Tau system. An analytical model of a momentum-driven bubble is shown to be consistent with our interpretation. We conclude that the bipolar outflow from DG Tau is intrinsically symmetric, and the observed asymmetries are due to environmental effects. This mechanism can potentially be used to explain the observed bipolar asymmetries in other YSO outflows.

\end{abstract}


\section{Introduction}\label{sec:intro}

Outflows are ubiquitous components of young stellar objects (YSOs). Solar-mass YSOs are capable of driving collimated bipolar outflows of atomic and molecular material to distances of $\sim 1\textrm{ pc}$ \citep[e.g.,][]{MRF07}. These outflows extract angular momentum from the star-disc system, allowing material from the circumstellar disc to accrete onto the central protostar. It is generally accepted that the outflows are launched magnetocentrifugally, either from the surface of the circumstellar disc \citep[MHD disc wind;][]{BP82,PN83}, or from reconnection points in the stellar magnetosphere \citep[the X-wind;][]{Se94}. Multiple launch mechanisms may act in concert to produce outflows with multiple velocity components \citep{Ae03,L03,FDC06,SLH07}. The launch region of these outflows is unresolvable with current telescopes. Therefore, detailed observational and theoretical studies of YSO outflows are necessary in order to determine the manner in which they are launched, and the physical conditions at their launching point(s).

Bipolar outflow asymmetry is a common occurrence in large-scale YSO outflows. These outflows, which are characterised by the presence of shock-excited Herbig--Haro (HH) objects, are often observed to be one-sided, with only a blueshifted, or approaching, outflow visible \citep[e.g.,][]{EM98,MR04,MRF07}. Such an asymmetry may be caused by the circumstellar disc obscuring the receding component of a symmetric bipolar outflow. Alternatively, the receding outflow may have entered the dense molecular cloud complex behind the YSO, obscuring it from observation \citep{MRF07}. However, radial velocity asymmetry is often seen in objects with observable bipolar HH outflows. For example, \citet{Hie94} found that, of 15 T Tauri stars with observed bipolar HH outflows, 8 showed bipolar velocity asymmetry between the blueshifted and redshifted outflows. The ratio of radial velocities between the opposing outflows in these objects is in the range 1.4--2.6. Further studies have found more asymmetric bipolar HH outflows, such as HH 30, which has a radial velocity ratio $\sim 2$ between the two outflows \citep{Ee12}. One-sided knot ejections have also been detected, such as that from the driving source of the HH 111 outflow \citep{GRL12}. Asymmetrical knot ejections and differing mass outflow rates between the two sides of the bipolar outflow from the Herbig Ae star HD 163296 have also been detected \citep{We06}. This evidence raises the question of whether the asymmetry is caused by environmental effects \citep{Hie94}, or is an intrinsic feature of the outflows, either due to disc conditions (such as warping) affecting the outflow launching \citep{GRL12}, or to other effects close to the launch point \citep{We06}.

With the advent of space-based telescopes such as the Hubble Space Telescope (HST), and the development of ground-based adaptive-optics systems, the large-scale outflows can be traced back to within a few hundred AU of the central protostar, and are observed as well-collimated `microjets'. These small-scale outflows provide an excellent laboratory for testing outflow launch models, as the outflow has yet to propagate to a distance where it interacts with the large-scale molecular cloud complex \citep[e.g.,][]{MRF07}. Therefore, the search for bipolar outflow asymmetry in these microjets is important in determining if the asymmetry is an intrinsic property of the outflows on all scales.

Velocity asymmetries were observed in the profiles of bipolar forbidden emission line (FEL) regions in several young stars \citep{HMS97}. The forbidden emission lines trace the presence of microjets \citep[e.g.,][]{BE99,Be02,Ce04,Ce07}. Further long-slit optical and near-IR spectroscopic observations have confirmed kinematic and/or physical bipolar asymmetries occur in the outflows of DG Tauri B \citep{Pe11} and FS Tauri B \citep{Lie12}. There are conflicting views on the cause of these asymmetries. \citet{Pe11} argue that the asymmetry in DG Tau B is due to an asymmetric ambient medium, based on the observation that only one side of the bipolar outflow is driving the ambient medium into a CO outflow. \citet{Lie12} argue for a bipolar outflow that is being driven at a different mass-loss rate on either side of the circumstellar disc, with the velocity difference between the two jets keeping a linear momentum balance, so there is no observable recoil. It is important to differentiate between the possible causes of bipolar outflow asymmetry in order to determine if it is an intrinsic or an environmental effect.

HST and adaptive optics also permits imaging and spectroimaging studies of YSO microjets \citep[e.g.,][]{Ke93,Le97}. Such studies have shown the presence of structural differences in the bipolar small-scale outflows from YSOs. The blueshifted collimated jet from the YSO HL Tauri is spatially coincident with an approximately axisymmetric bubble-like structure \citep{Te07}. Bipolar asymmetries in jet collimation are observed in the YSOs RW Aurigae \citep{YMe09} and DG Tau B \citep{Pe11}. Such studies have shown that structural bipolar outflow asymmetries are common in YSOs on the microjet scale \citep{Pe11}.

Another example of bipolar asymmetry in T Tauri star microjets is the transitional Class I/Class II YSO DG Tauri. One of the most actively-accreting T Tauri stars, DG Tau has been used as a laboratory in searches for jet rotation \citep{B02,Pe04,Ce07}, jet knot generation \citep{LFCD00,Re12} and links between jet and disc properties \citep{Tes02}. DG Tau drives the blueshifted HH 158 \citep{MF83} and HH 702 \citep{Se03,MR04} outflows, the latter extending to a distance of $\sim 0.5\textrm{ pc}$ from the protostar. There is no known large-scale redshifted HH outflow associated with DG Tau. A bipolar microjet-scale outflow is present, and exhibits velocity asymmetry between the approaching and receding flows \citep{He94,Le97}. This was originally detected through long-slit spectroscopy of optical forbidden emission lines from the outflows \citep{Hie94}. The receding outflow was first imaged by \citet{Le97} using spectroimaging of [O I] 6300 \AA\ emission. They determined a radial velocity ratio of 1.4 between the approaching and receding outflows.

More recently, \citet{A-Ae11} and \citet{MCW13a} (hereafter referred to as Paper I) detected structural and kinematic differences in the microjet-scale approaching and receding outflows of DG Tau, using spectroimaging data of spatially extended [Fe II] 1.644 $\mu$m line emission. The approaching outflow shows the classical YSO microjet morphology of a central, well-collimated, high-velocity jet with deprojected velocity $\sim 215\textrm{--}315\kms$. The jet is dominated by both stationary and moving shock-excited `knots' of emission. This jet is `nested' within a region of lower-velocity emission, which may be excited by the formation of a turbulent entrainment layer around the jet \citep[Paper I;][]{Pe03b}. A wide-angle approaching molecular wind is observed in H$_2$ 1-0 S(1) 2.128 $\mu$m line emission \citep[Paper I;][]{Be08}, providing a supply of material for the jet to entrain. On the other hand, the redshifted outflow shows no evidence of any jet-like components, and instead forms a large bubble-like structure. This was interpreted by \citet{A-Ae11} as being the counterpart `magnetic bubble' \citep{Cie09} to a similar structure they claimed $\sim 1\farcs 2$ from the central star in the approaching outflow channel. However, analysis in Paper I showed that they appeared to be interpreting the low-velocity entrainment component in that region as part of the central jet. We concluded that an approaching bubble structure does not exist (Paper I). Therefore, the nature and cause of the bipolar outflow asymmetry in DG Tau remains an open question.

We investigate the bipolar asymmetry in the microjet-scale DG Tau outflows below, and conclude that environmental effects hamper the propagation of one side of an approximately symmetric bipolar outflow. We proceed as follows. In \S\ref{sec:obs}, we outline our multi-epoch NIFS observations and our data reduction procedure. \S\ref{sec:analysis} details our methods of analysing the data. In \S\ref{sec:bubble}, we argue that the structure is a stationary bubble with an internal velocity field describing expansion of gas towards the bubble walls, based on comparisons of the observed velocity structure to kinetic models. \S\ref{sec:bubbleAGN} outlines the results of simulations of bubbles driven by impeded active galactic nuclei (AGN) jets, and links this work to the morphology observed in the DG Tau receding outflow. We propose that the bubble in the DG Tau receding outflow is the result of a receding counterjet being obstructed by clumpy ambient material in the extended envelope around DG Tau \citep{KKS96a}. The receding outflow is currently in the momentum-driven bubble phase, similar to the simulations of radio galaxies by \citet{SB07} and \citet{WB11}. We construct an analytical model of an expanding jet momentum-driven bubble in \S\ref{sec:model}, and find that it predicts physical parameters consistent with those observed in the DG Tau bubble and the extended CO envelope around the system. Finally, in \S\ref{sec:D}, we discuss the impact of our results on the interpretation of bipolar outflow asymmetry in other YSOs, as well as the implications of episodic variability in YSOs on our model. We summarise our conclusions in \S\ref{sec:concl}.

\section{Observations \& Data Reduction}\label{sec:obs}

A brief outline of the data reduction techniques used is given here. For a more detailed description, the reader is referred to Paper I.

\begin{table*}
\caption{Observing parameters of DG Tau $H$-Band Observations, 2005--2009}\label{tab:obsparams}
\begin{tabular}{ccccccccc}
\hline
\multicolumn{1}{c}{Date} & \multicolumn{1}{c}{Epoch} & \multicolumn{1}{c}{No.~of} & \multicolumn{3}{c}{Telluric standard star}  & & \multicolumn{2}{c}{Continuum image FWHM} \\
\cline{4-6} \cline{8-9}
\multicolumn{1}{c}{} & \multicolumn{1}{c}{} & \multicolumn{1}{c}{on-source} & \multicolumn{1}{c}{Star} & \multicolumn{1}{c}{2MASS $H$} & \multicolumn{1}{c}{Blackbody} & & \multicolumn{1}{c}{Telluric} & \multicolumn{1}{c}{DG Tau} \\
\multicolumn{1}{c}{} & \multicolumn{1}{c}{} & \multicolumn{1}{c}{exposures\textsuperscript{a}} & \multicolumn{1}{c}{} & \multicolumn{1}{c}{(mag)} & \multicolumn{1}{c}{temp.~(K)} & & \multicolumn{1}{c}{standard\textsuperscript{b}} & \multicolumn{1}{c}{} \\
\hline
2005 Nov 11 & 2005.87 & 11 & HIP25736 & 7.795 & 7000 & & $0\farcs 11$ & $0\farcs 14$ \\
2006 Dec 24 & 2006.98 &  9 & HIP25736 & 7.795 & 7000 & & $0\farcs 11$ & $0\farcs 12$ \\
2009 Nov 08 & 2009.88 &  6 & HIP26225 & 7.438 & 9400 & & $0\farcs 09$ & $0\farcs 18$ \\
\hline
\end{tabular}
\begin{flushleft}
\textsuperscript{a}All on-source exposures were 600 s.\\
\textsuperscript{b}Used to estimate the AO seeing achieved.
\end{flushleft}
\end{table*}

Observations of the DG Tau system were obtained on 2005 Oct 26 UT in the $K$-band \citep{Be08}, and on 2005 Nov 12 UT in the $H$-band (Paper I), with the Near-infrared Integral Field Spectrograph (NIFS) on the Gemini North telescope, Mauna Kea, Hawaii. Further $H$-band observations were taken on 2006 Dec 24 and 2009 Nov 08. Observing parameters for each epoch of $H$-band data are detailed in Table \ref{tab:obsparams}. Data were recorded with the ALTAIR adaptive-optics system in natural guide star mode, utilising DG Tau itself as the reference star. NIFS is an image-slicing type integral-field spectrograph, splitting a $3\arcsec\times 3\arcsec$ field into 29 slitlets, resulting in $0\farcs 103\times 0\farcs 045$ spaxels. A spatial resolution of $\sim 0\farcs 1$ was achieved in the $H$-band data, based on the FWHM of a standard star observed immediately after the DG Tau observations. These standard star observations allow for telluric correction and flux calibration. Calibration was based on the 2MASS magnitude of the standard star, and a shape derived from a blackbody function fit to the 2MASS $J$-$K$ colour of the standard star. A spatial resolution of $\sim 0\farcs 1$ was achieved in the $K$-band observations \citep{Be08}. A $0\farcs 2$-diameter partially-transmissive occulting disc was placed over the star during the $H$-band observations to increase sensitivity to extended emission.

Data reduction was performed using the Gemini {\sc IRAF} package as follows. An average dark frame was subtracted from each object frame and averaged sky frame. The dark-subtracted average sky frame was then subtracted from the dark-subtracted object frame. A flat-field correction was applied to each slitlet by dividing by a normalised flat-field frame. Bad pixels were identified using the flat-field and dark frames, and were corrected via 2D interpolation. Individual 2D spectra for each slitlet were transformed to a rectilinear coordinate grid using arc and spatial calibration frames, and then stacked in the second spatial direction to form a 3D data cube. All spectra were transformed to a common wavelength scale at this point.

Data cubes from each object exposure were corrected for telluric absorption by division with a normalised 1D spectrum extracted from the observations of a telluric standard star. Hydrogen absorption lines intrinsic to the A0 standard star were removed using Gaussian fits to those lines. Flux calibration was achieved using a large-aperture 1D spectrum of the same standard star, which was also corrected for telluric absoprtion. The finalised object frames were spatially registered using the continuum position of DG Tau, and median-combined to produce the final data cube. The final cube for each epoch was then spatially registered using the position of the central star to allow for comparison of the extended emission structure.

Stellar subtraction was performed using custom {\sc Python} routines. For $H$-band data cubes, two $0\farcs 25$ diameter apertures were formed, centered at opposing positions $0\farcs 5$ from the central star perpendicular to the outflow direction. A stellar spectrum was extracted from each aperture and averaged. For each spaxel in the cube, this stellar spectrum was scaled to match the flux observed adjacent to the spectral region of interest for the line being investigated, and subtracted from the spaxel spectrum. $K$-band stellar subtraction was performed by forming a pair of continuum images adjacent to the spectral region of interest around the line being investigated. These images were averaged, and then subtracted from each wavelength plane of the data cube. Full details of the data reduction and stellar spectrum subtraction procedures used may be found in \citet{Be08} ($K$-band) and Paper I ($H$-band). We concentrate on the 2005 epoch data in this paper; unless explicitly stated, all data are from that epoch.

\begin{figure*}
\centering
\includegraphics[width=\textwidth]{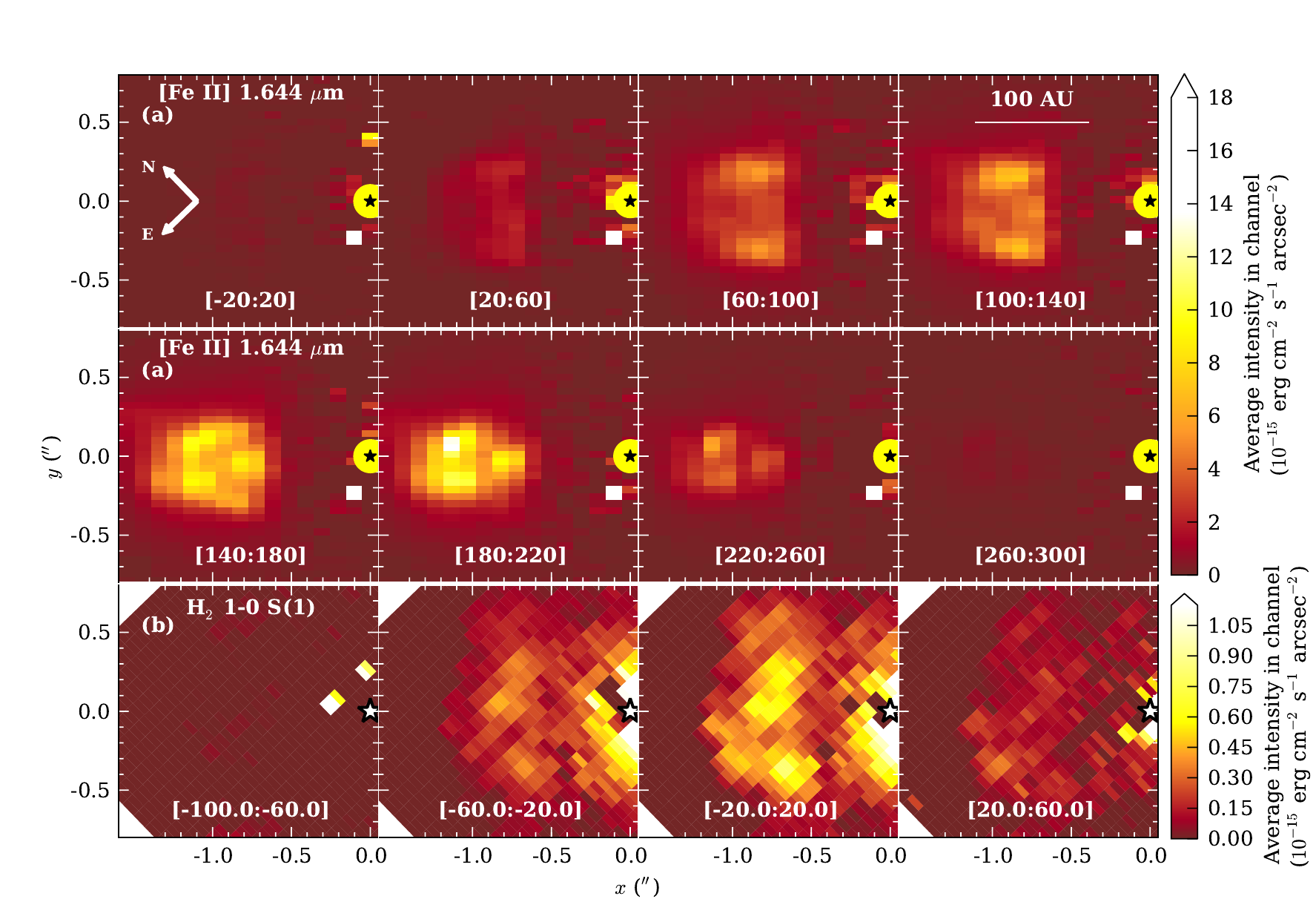}
\caption{Channel maps of emission from the receding DG Tau outflow --- 2005 epoch. Panels show images of (a) the extended [Fe II] 1.644 $\mu$m line emission, and (b) the extended H$_2$ 1-0 S(1) 2.1218 $\mu$m line emission from the receding DG Tau outflow, binned into 40 km s$^{-1}$-wide slices. The velocity range of each slice is shown in white in$\kms$. The intensity values quoted are the average intensity in each channel over the velocity range. The black star corresponds to the position of the central star, DG Tau, and the yellow circle in (a) indicates the position and size of the occulting disc used during the $H$-band observations. The physical scale is indicated at the top-right. This scale does not account for the inclination of the outflow axis, $x$, to the line of sight.}\label{fig:redchannels}
\end{figure*}

The receding outflow from DG Tau is shown in Fig.~\ref{fig:redchannels}, as seen in [Fe II] 1.644 $\mu$m line emission (Fig.~\ref{fig:redchannels}(a)) and H$_2$ 1-0 S(1) 2.1218 $\mu$m line emission (Fig.~\ref{fig:redchannels}(b)). In this Figure, as with all following figures, the large-scale HH 158 outflow axis is labelled as $x$, and the axis transverse to this as $y$. The extended $H$-band emission is dominated by [Fe II] 1.644 $\mu$m line emission, and takes the form of a bubble-like structure. The `apex' of this structure is $\sim 1\farcs 3$ along the outflow axis from the central star. No emission is observed closer than $\sim 0\farcs 7$ to the central star, due to obscuration by the circumstellar disc. This obscuration provides a measure of the extent of the disc \citep{A-Ae11}. The lateral width of the structure at its closest to the central star, which we shall refer to as the `base' of the structure, is $\sim 0\farcs 7\approx 98\textrm{ AU}$ at the distance to DG Tau \citep[140 pc;][]{E78}. The lowest line velocities from the structure occur at this widest point; the highest line velocities occur at both the apex of the structure, and at an emission enhancement $0\farcs 84\pm 0\farcs 03$ along the outflow axis from the central star.

The H$_2$ 1-0 S(1) 2.1218 $\mu$m line emission from the region of the receding outflow is strongest at positions coincident with the base of the bubble-like structure seen in [Fe II] 1.644 $\mu$m line emission \citep[Fig.~\ref{fig:redchannels}(b);][]{Be08,A-Ae14}. This `patchy' emission has no clear structure. The line emission is concentrated about a velocity of $0\kms$. \citet{Be08} analysed the level populations and line ratios in the observed H$_2$ emission, and determined that it is likely being excited by collisional J- or C-type shocks. UV and X-ray fluorescence were previously ruled out by \citet{Te04}, based on the large mass and momentum fluxes resulting from their calculations.

\section{Data Analysis}\label{sec:analysis}

\subsection{Spectral Gaussian Fitting}\label{sec:obs-fitting}

Upon visual inspection, some regions of the receding outflow appear to exhibit multiple components in [Fe II] 1.644 $\mu$m line emission. Multi-component spectral Gaussian fitting was used to characterise the presence and nature of these components. Both one- and two-component Gaussian fits were made to the [Fe II] 1.644 $\mu$m emission line in each spaxel, and an $F$-test \citep[Paper I, appendix A;][]{BR92} was used to determine the statistically appropriate number of components to retain in the final fit \citep{Wee07}. It has been noted that the $F$-test is formally not the correct test to use in this situation \citep{Pre02}. However, lacking a statistically correct specific alternative that may be applied to the volume of spectra presented here, we proceed using the $F$-test \citep[Paper I;][]{Wee12}. It was found that a two-component fit could not be consistently applied across the entire receding outflow. This occurs because the two apparent components are too convolved to form a statistically significant fit with our spectral resolution. Alternatively, there may be an underlying `continuum' of emission velocity components at some spatial positions. This effect is spread evenly across the receding outflow structure. Therefore, Gaussian line fits were restricted to one component at all spatial positions. This is in contrast to the approaching outflow, which shows the clear presence of two [Fe II] 1.644 $\mu$m emission-line components at all spatial positions (Paper I).

\begin{figure}
\centering
\includegraphics[width=0.5\textwidth]{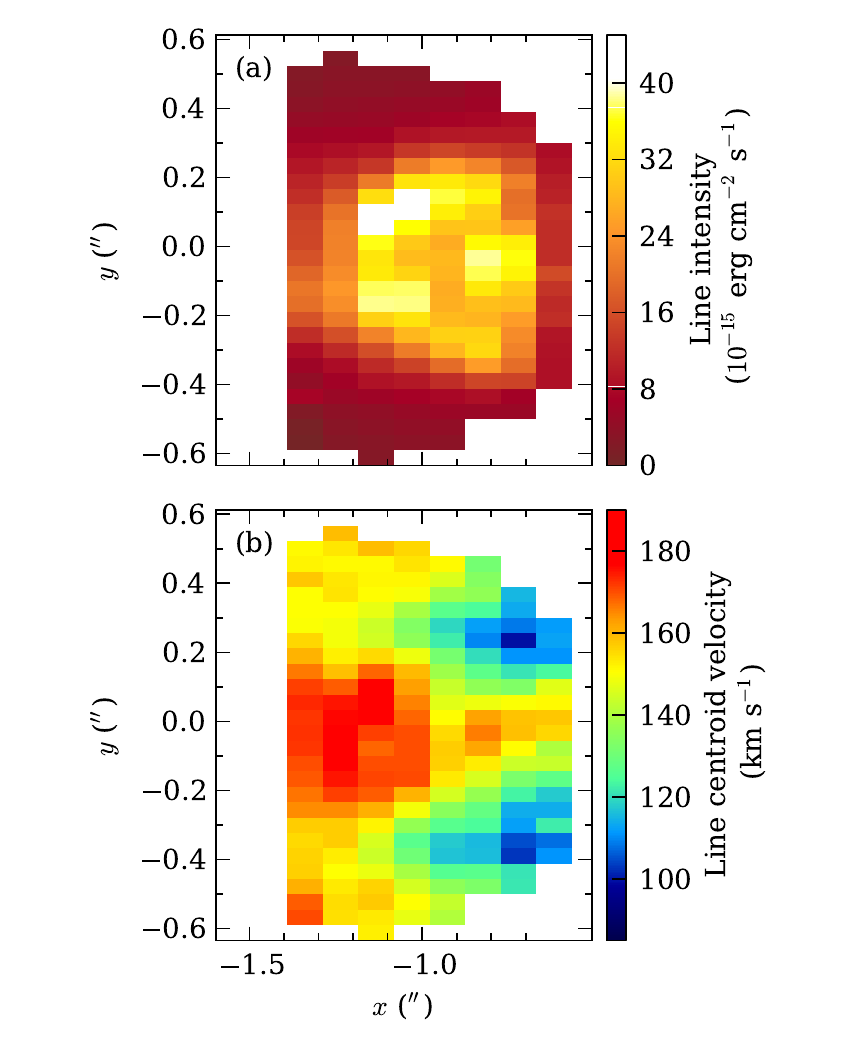}
\caption{Fitted [Fe II] 1.644 $\mu$m emission-line component for the DG Tau receding outflow. Panel (a) shows the fitted line intensity and panel (b) displays the fitted line velocity of the receding outflow, based on a single-component Gaussian fit. The fitted line velocity has been corrected for the systemic velocity of the central star, as determined by Gaussian fitting to observed $H$-band photospheric absorption features. This Figure was presented previously in Paper I.}\label{fig:redvelfit}
\end{figure}

The results of the spectral fitting procedure are shown in Fig.~\ref{fig:redvelfit}. Line velocities were corrected for the systemic velocity of the central star, as determined by Gaussian fits to multiple photospheric absorption lines observed in the $H$-band stellar spectrum (Paper I, fig.~1 therein). Comparison of Fig.~\ref{fig:redvelfit}(a) to the channel maps of [Fe II] 1.644 $\mu$m line emission (Fig.~\ref{fig:redchannels}(a)) shows that the fitting procedure accurately replicates the appearance of the receding outflow.

The velocity structure observed in the receding outflow is in agreement with \citet{A-Ae11}. The largest receding line-of-sight velocities of $\sim 160\textrm{--}180\kms$ are observed at the apex of the structure (Figs.~\ref{fig:redchannels}, \ref{fig:redvelfit}(b)), as well as along the outflow axis. Line velocities on the edges of the structure decrease with decreasing distance from the central star. The lowest line velocities, $\sim 80\kms$, are located at the base of the observable structure.

\subsection{Electron Density}\label{sec:obs-density}

The near-infrared lines of [Fe II] arise from low-lying energy levels, and are useful tracers of electron number density, $n_\textrm{e}$. \citet{PZ93} showed that the intensity ratio of the [Fe II] lines at wavelengths 1.533 $\mu$m and 1.644 $\mu$m are a diagnostic of electron number density in the range $n_\textrm{e}\sim 10^2\textrm{--}10^6\textrm{ cm}^{-3}$, for electron temperatures, $T_\textrm{e}$, in the range $T_\textrm{e}\sim 3000\textrm{--}20000\textrm{ K}$. \citet{Pes03} have accurately computed the relation between the line ratio, $F_{1.533}/F_{1.644}$, and electron density for a 16-level Fe$^+$ model.

\begin{figure}
\centering
\includegraphics[width=0.5\textwidth]{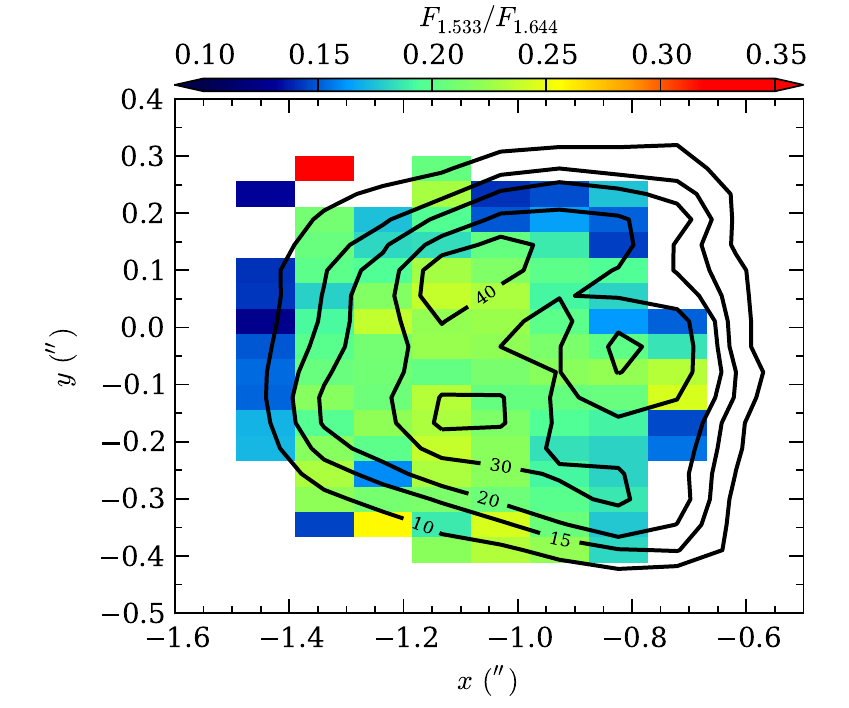}
\caption{Ratio of [Fe II] 1.533 $\mu$m to [Fe II] 1.644 $\mu$m emission-line intensity from the receding DG Tau outflow. Spaxels have been masked where a threshold signal-to-noise ratio of 10 in the computed ratio has not been met, or where the ratio is in the saturation limit for determining electron density \citep[$F_{1.533}/F_{1.644}\gtrsim 0.45$;][]{Pes03}. Line fluxes are determined by integration of the raw stellar-subtracted spaxel spectra about the line wavelengths, over the velocity range $0$ to $340\kms$. Contours of [Fe II] 1.644 $\mu$m line emission, integrated over the same velocity range, are overlaid in black. Contours are labelled in units of $10^{-15}\textrm{ erg cm}^{-2}\textrm{ s}^{-1}\textrm{ arcsec}^{-2}$.}\label{fig:reddens}
\end{figure}

Fig.~\ref{fig:reddens} shows the flux-ratio map derived from our data for the receding DG Tau outflow. Integrated line fluxes were determined by the integration of the stellar-subtracted spectrum of each spaxel over the velocity range $0$ to $340\kms$ about the two line wavelengths. Spaxels were excluded from this calculation where a signal-to-noise ratio of $10$ in the computed line ratio was not achieved, meaning line ratios could only be determined where the weaker 1.533 $\mu$m emission line could be detected with adequate signal-to-noise ratio. This criterion produces density information over a region comparable to the observed redshifted [Fe II] 1.644 $\mu$m line emission (Fig.~\ref{fig:redchannels}).

The $F_{1.533}/F_{1.644}$ ratio is approximately constant across the region of observable redshifted emission (Fig.~\ref{fig:reddens}). The relationship between the line ratio and electron density is only weakly dependent on electron temperature, especially in the range $0.1 \lesssim F_{1.533}/F_{1.644}\lesssim 0.3$ \citep[][fig.~2(b) therein]{Pes03}, so we can comment on the approximate electron density of the receding outflow without knowledge of the electron temperature. We conclude that the electron density of the receding outflow is of order $10^{4}\textrm{ cm}^{-3}$. It may rise to $\sim 10^{4.5}\textrm{ cm}^{-3}$ in the regions of strongest [Fe II] 1.644 $\mu$m emission. However, these variations in $F_{1.533}/F_{1.644}$ are barely larger than the uncertainties in their calculation.

\subsection{Time-Evolution of the Receding Outflow}\label{sec:analysis-time}

\begin{sidewaysfigure*}
\centering
\includegraphics[width=\textwidth]{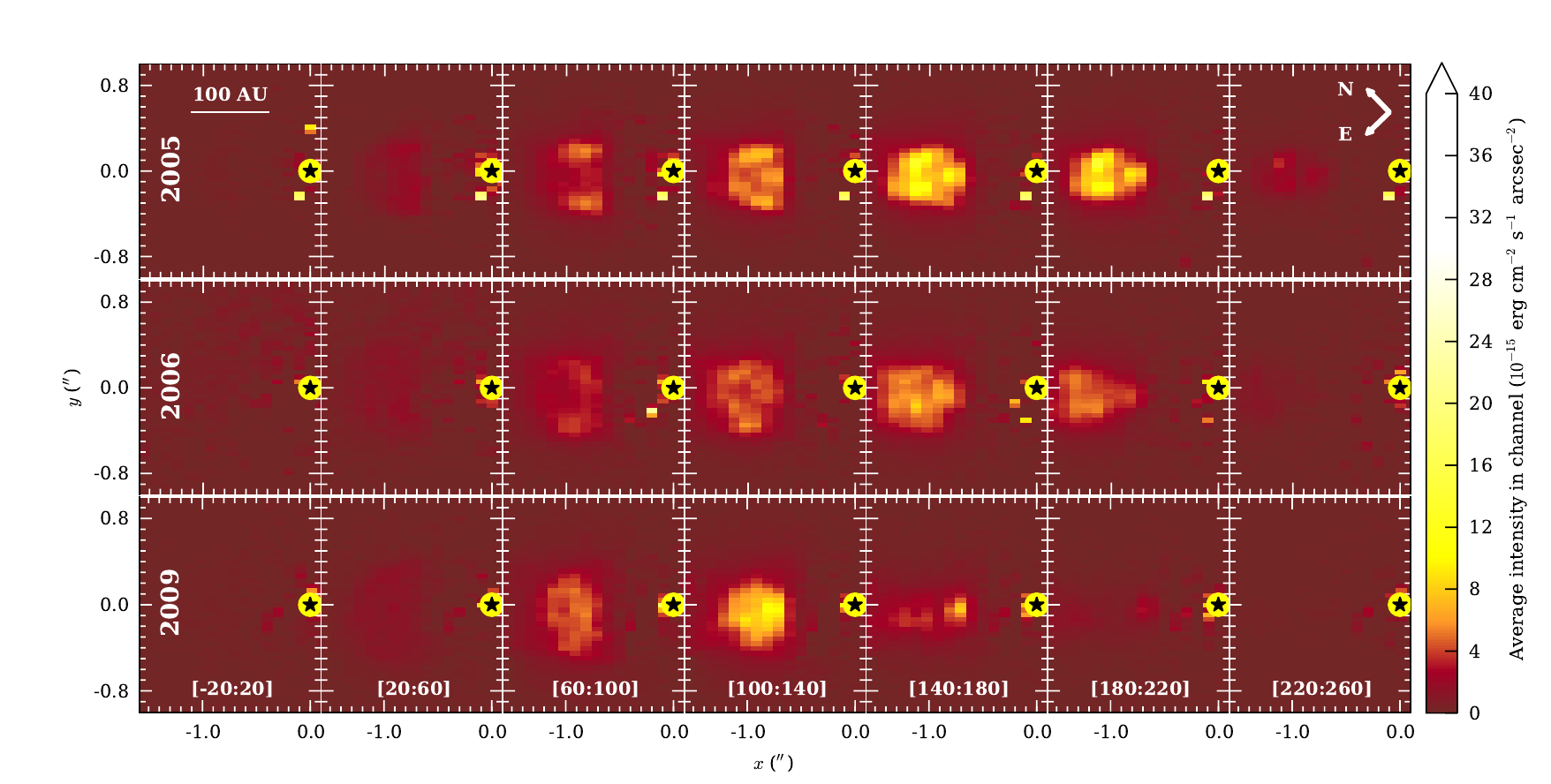}
\caption{Channel maps of the DG Tau outflow for the 2005, 2006 and 2009 observing epochs. Panels show images of the extended [Fe II] 1.644 $\mu$m line emission around DG Tau, binned into $40\kms$-wide slices. The velocity range of each slice is shown at the bottom of each slice in$\kms$. The intensity values quoted are the average intensity in each channel over the $40\kms$ velocity range. The black star corresponds to the position of the central star, DG Tau, and the yellow circle indicates the position and size of the $0\farcs 2$ diameter occulting disk.}
\label{fig:redslices050609}
\rule{0pt}{0.75\textwidth}
\end{sidewaysfigure*}

\begin{figure}
\centering
\includegraphics[width=\columnwidth]{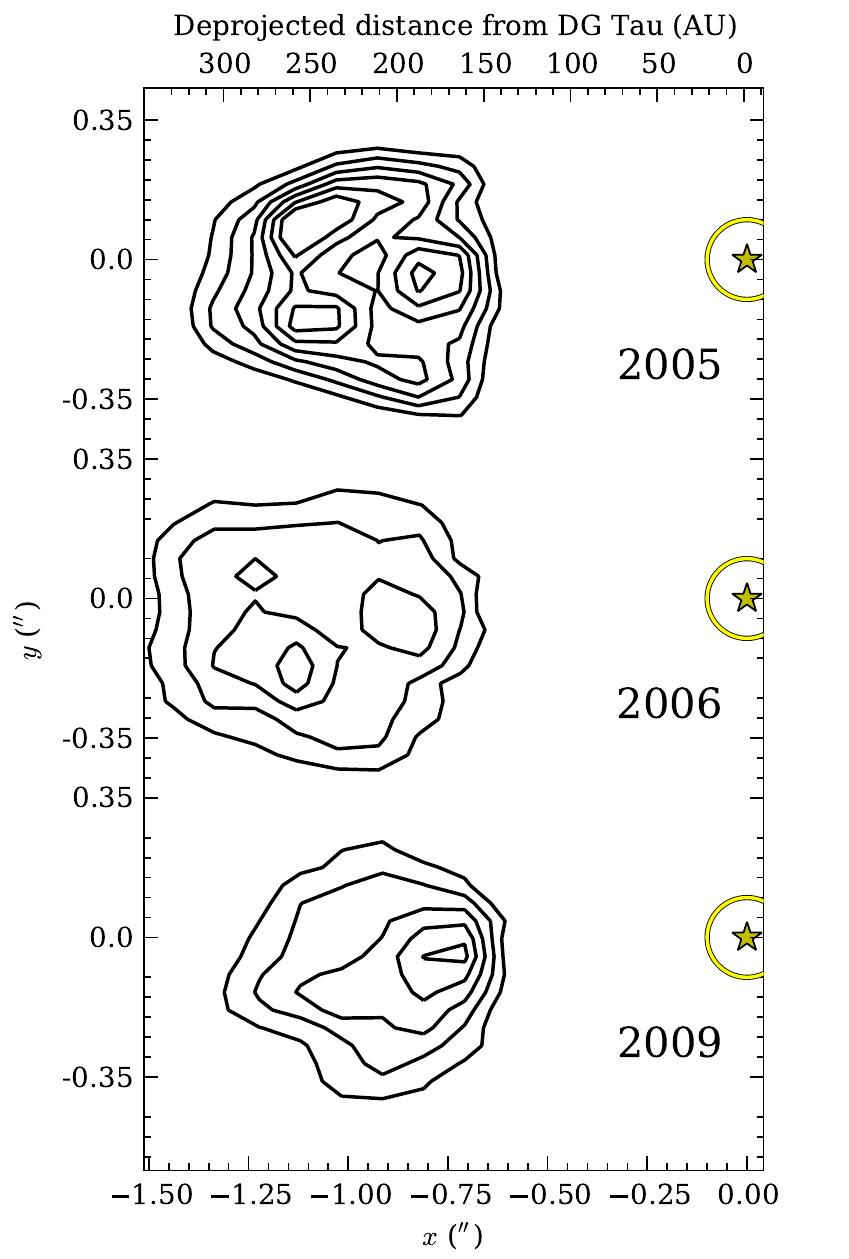}
\caption{[Fe II] 1.644 $\mu$m contour image of the receding outflow from DG Tau, formed over the velocity range $0$ to $340\kms$, for observing epochs 2005--2009. Contours are drawn at levels of $[15,20,\ldots,40]\times 10^{-15}\textrm{ erg cm}^{-2}\textrm{ s}^{-1}\textrm{ arcsec}^{-2}$. The location of the central star, and the position and size of the occulting disc used during the observations, are shown as a yellow star and circle, respectively.}
\label{fig:redcontours}
\end{figure}

The time-evolution of the receding outflow over the period 2005--2009 structure is shown in Figs.~\ref{fig:redslices050609} and \ref{fig:redcontours}. The most remarkable feature is the apparent stability of the bubble structure between the 2005 and 2006 observing epochs, 1.11 yr apart. The velocity structure is particularly stable over this period (Fig.~\ref{fig:redslices050609}). Furthermore, the faint boundary of low-velocity emission remains constant across all epochs (Fig.~\ref{fig:redslices050609}, $[20:60]\kms$ panel). The greatest change occurs between the 2006 and 2009 observing epochs, when the `apex' of the structure is no longer visible. The `apex' is formed of the highest-velocity emission (Fig.~\ref{fig:redslices050609}).

We investigate the stability of the receding outflow structure between the 2005 and 2006 observing epochs more carefully. The `base' of the structure remains in the same position in all observing epochs, $\sim 0\farcs 7$ from the central star (Fig.~\ref{fig:redcontours}). Based on the extent of the $15\times 10^{-15}\textrm{ erg cm}^{-2}\textrm{ s}^{-1}\textrm{ arcsec}^{-2}$ contour in Fig.~\ref{fig:redcontours}, the `apex' of the structure appears to have moved $\sim 0\farcs 1$ in 1.1 yr, corresponding to a deprojected\footnote{We assume an inclination between the outflow axis and the line of sight of $38^\circ$ \citep{EM98}.} velocity of $\sim 95\kms$. This apparent motion occurs on the scale of a single NIFS pixel, and the scale of the AO-corrected seeing (Table \ref{tab:obsparams}). Furthermore, the small discrepancy ($0\farcs 03$) between the position of knot A of the approaching outflow in the 2006 observing epoch, and in the 2005 and 2009 epochs (Paper I, table 2 therein), suggests there may be a larger spatial registration uncertainty in the 2006 epoch data. This may reduce the actual bubble `apex' proper motion to $\sim 70\kms$. The main difference between the receding outflow structure between the 2005 and 2006 epochs is in the highest-velocity emission (Fig.~\ref{fig:redslices050609}, panels $[140:180]$ and $[180:220]\kms$). The stationarity of (a) the lower-velocity emission and (b) the `base' of the structure leads us to conclude that the receding outflow structure is a predominantly stationary feature. We now proceed to consider a possible model for the formation of such a structure.

\section{Receding Outflow as a Bubble}\label{sec:bubble}

The nature of the DG Tau receding outflow structure has yet to be adequately determined. Such structures are not unique to DG Tau, having been observed in other YSOs \citep[e.g., HL Tauri,][]{Te07}. \citet{A-Ae11} interpreted the receding outflow structure in DG Tau to be the counterpart of a similar faint bubble they reported at and beyond $1\farcs 2$ from the central star in the approaching outflow. We do not observe this approaching bubble (Paper I). We believe that \citet{A-Ae11} were interpreting the blueshifted intermediate-velocity component (IVC) at that location as being a separate structure to the rest of the IVC. The IVC accelerates with distance from the central star, so that at distances $\gtrsim 1\farcs 2$ from the central star, the IVC approaches the same velocity that the high-velocity jet exhibits closer to the star. This could lead to the presumption that this material is of the same origin as the jet, particularly if a simple velocity cut is used to differentiate between components, as is the case in \citet{A-Ae11}. We conclude that there is no approaching outflow counterpart to the receding outflow structure (Paper I).

Ignoring for the moment the question of how DG Tau produces two markedly different structures on two sides of a bipolar outflow (\S\ref{sec:bubbleAGN}), one potentially appealing interpretation for the presence of a bubble-like structure in the receding outflow is a bow shock. Bow shocks are the predominant components of the large-scale Herbig-Haro chains observed in many protostellar outflows on larger scales. On smaller scales, moving microjet knots (c.f.~Paper I) have been observed to have a mini-bow shock morphology when observed at high angular resolution \citep[e.g., HH 34,][]{Ree02}. However, all of these structures have been observed to possess significant proper motion, of the order of the jet propagation velocity \citep{Ree02,He05,Rae12}. This is not observed for the receding outflow structure in DG Tau. Furthermore, if the receding outflow were forming a bow shock, in the absence of significant receding proper motion it would exhibit significantly mixed redshifted \emph{and blueshifted} emission. This would be the result of the backflow of material that is ejected orthogonal to the jet axis at the bow shock head \citep{LF96}. We observe only redshifted emission on one side of the DG Tau system, and only blueshifted emission on the other (Paper I, fig.~2 therein). Therefore, we exclude the presence of a stationary bow shock-type feature in the DG Tau receding outflow.

\begin{figure}
\centering
\includegraphics[width=0.5\textwidth]{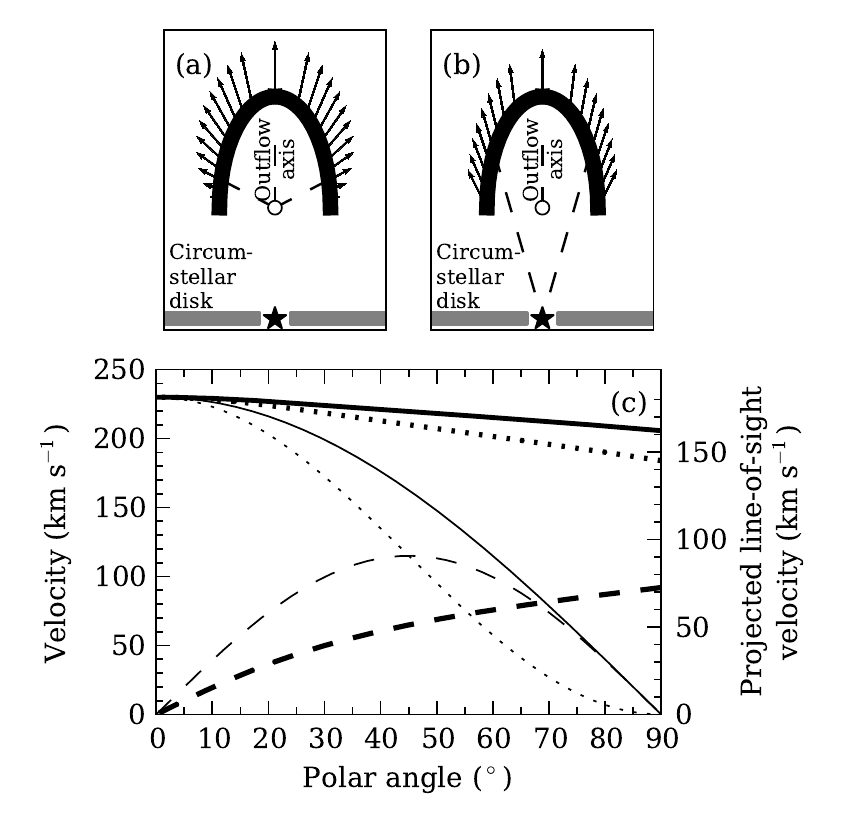}
\caption{Kinetic models of expanding bubbles. The top panels show velocity vectors for snapshots of hemi-ellipsoidal bubbles with velocity fields expanding (a) away from the centre of the bubble, and (b) away from the location of the central star. Note that the vectors represent the velocity field of the dispersed, shocked jet material interior to the bubble walls, which is the source of the emission; the vectors are placed on the outside of the bubble for clarity only. The central star, denoted by the star symbol, is at the same distance from the bubble centre, denoted by the small circle, as the bubble centre is from the bubble apex. The dashed lines show the alignment of the velocity vectors in each model. The bubble major axis is aligned to the outflow axis. The polar angle is defined as the angle from the outflow axis, measured at the bubble centre. Velocities are modulated by the polar angle such that the total gas expansion velocity reduces to zero at the same height above the notional circumstellar disc (grey regions) as the point from which the internal velocity field expands. The star and circumstellar disc are not included in the models. (c) Velocity profiles of the models. Profiles for the internal velocity field expanding away from the bubble centre (panel (a)) are drawn in thin lines; profiles for the bubble expanding away from the central star (panel (b)) are drawn in thick lines. The dashed lines show radial velocity profiles, the dotted lines show vertical velocity profiles, and the solid lines show total velocity profiles.}\label{fig:bubblecartoon}
\end{figure}

\begin{figure*}
\centering
\includegraphics[width=\textwidth]{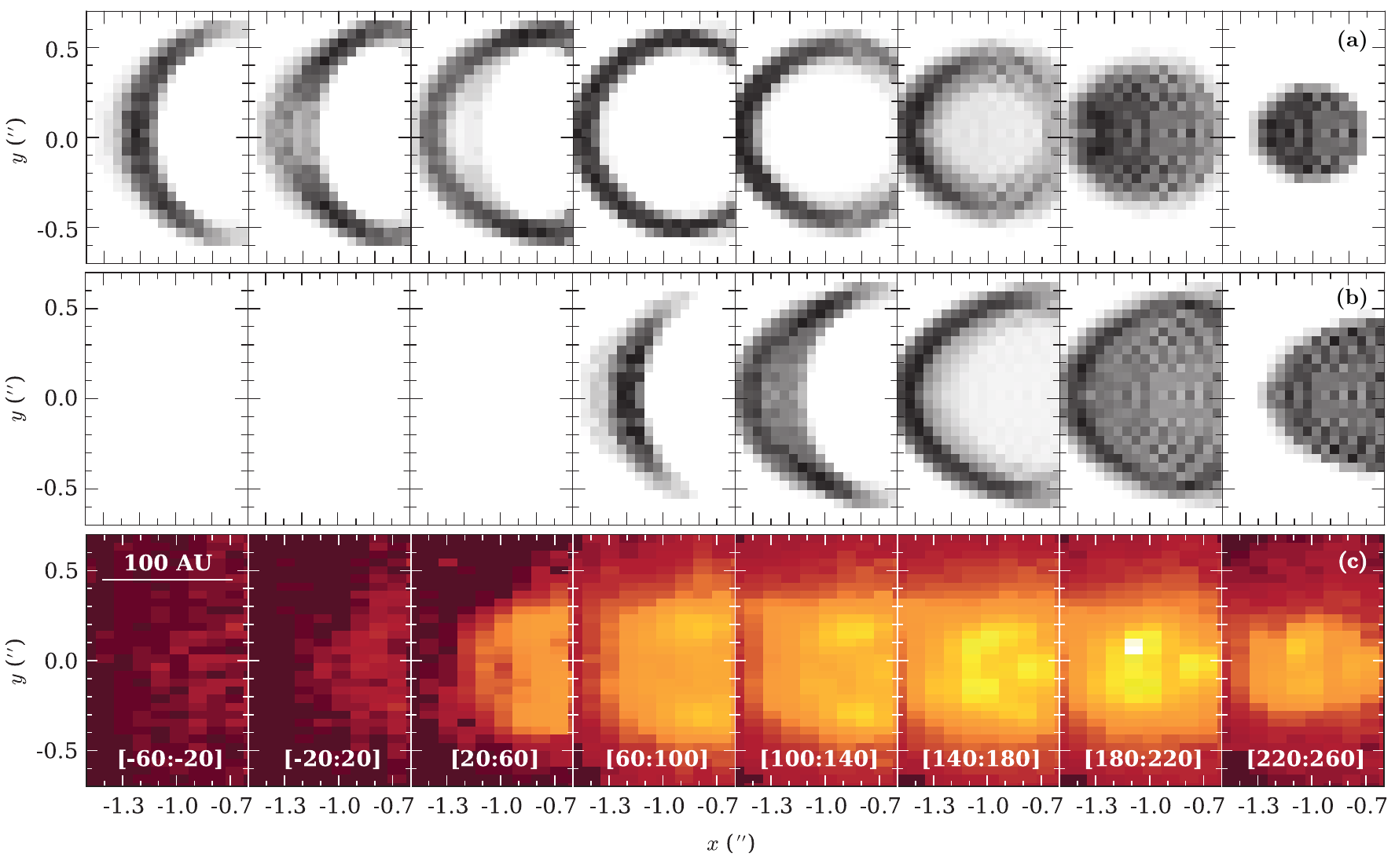}
\caption{Comparison of simulated IFU data of two static, radially symmetric hemi-ellipsoidal bubble models with an internal distribution of emitting material with an expanding velocity field. (a) Bubble model where gas expansion velocity vectors point away from the centre of the bubble. (b) Bubble model where gas expansion velocity vectors point away from the central star. In both cases, the major axis of the bubble is coincident with the outflow axis, and the bubble base is positioned $0\farcs 70$ from the central star. The emitting regions of the bubble have a fixed thickness of 15 AU. Maximum gas expansion velocities are set to $230\kms$. Gas expansion velocities peak at the `apex' of the bubble stucture, and decrease to $0\kms$ at the same height above the circumstellar disc as the point from which the bubble expands. Simulated data are binned into velocity slices based on a model IFU with $0\farcs 05\times 0\farcs 05$ spatial pixels and $21\kms$ spectral pixels. Intensity scaling between each velocity slice is different. (c) Channel maps of the receding bubble in DG Tau, as observed in [Fe II] 1.644 $\mu$m line emission. Intensity scaling is the same between individual panels.}\label{fig:bubblecompare}
\end{figure*}

We investigated the velocity structure of the [Fe II] 1.644 $\mu$m line emission from the receding outflow using kinetic models as follows. The models are constructed on a three-dimensional Cartesian grid using the {\sc Fortran} programming language, and represent snapshots of static bubbles, with a distribution of expanding, emitting material interior to the bubble walls. In the first model (Fig.~\ref{fig:bubblecartoon}(a)), the velocity field expands away from the intersection between the large-scale outflow axis and the base of the structure, which we call the bubble centre. In the second model (Fig.~\ref{fig:bubblecartoon}(b)), the velocity field expands away from the position of the central star. The central star is assumed to be located on the outflow axis, at the same distance from the bubble centre as the bubble apex. However, the star is not included in the models. Each bubble has a fixed emission region thickness of 15 AU interior to the bubble walls, within which every volume element radiates with the same emissivity. Volume elements interior to these regions do not radiate. The velocity of gas in the emitting regions is modulated by the angle from the outflow axis, such that the expansion velocity decreases to zero at the same height above the central star as the point from which the bubble expands. This was necessary in order to limit the amount of blueshifted emission predicted by the models. The height (160 AU), elongation (height-to-width ratio $\sim 2$) and position relative to the central star (160 AU) of the bubble were chosen to approximately match the observed dimensions of the DG Tau bubble (Fig.~\ref{fig:redchannels}). The maximum expansion velocity is set to $230\kms$, matching the highest deprojected line velocity observed in the bubble-like structure (Fig.~\ref{fig:redvelfit}(b)). The models account for the inclination of the DG Tau jet-disc system to the line-of-sight, and generate a simulated channel map of the emission from the bubble, based on an IFU with $0\farcs 05\times 0\farcs 05$ spaxels and $21\kms$ spectral pixels. The simulated channel maps generated from these models are shown in Fig.~\ref{fig:bubblecompare}.

The redshifted outflow of DG Tau is well-described phenomenologically by a bubble with an internal velocity field describing expansion towards the bubble walls. Our models produce simulated channel maps in reasonable agreement with the observational data. However, the model of the bubble material expanding away from the position of the central star (Figs.~\ref{fig:bubblecartoon}(b), \ref{fig:bubblecompare}(b)) produces less blueshifted emission than the bubble material expanding away from the bubble centre (Figs.~\ref{fig:bubblecartoon}(a), \ref{fig:bubblecompare}(a)). Therefore, the model where the bubble material expands away from the central star is in better agreement with observations (Figs.~\ref{fig:redchannels}(a), \ref{fig:bubblecompare}(c)). This suggests that the location where energy and/or momentum (\S\ref{sec:model-applic}) is being dispersed to drive the bubble expansion is closer to the central star than the centre of the bubble. We cannot definitively determine the location of this driving centre because of obscuration by the circumstellar disc (although, see \S\ref{sec:bubbleAGN-jet}). However, it is clear that a simple kinetic bubble model provides a good approximation to the channel maps of the receding outflow structure.

\section{Origins of Asymmetric Outflows from AGN modeling}\label{sec:bubbleAGN}

We have determined that the redshifted emission from the DG Tau outflow is well-described by a model of a stationary bubble with an internal distribution of expanding, emitting material. This raises the question as to why a bubble has formed on only one side of the DG Tau star-disc system, whereas the opposing outflow has been propagating as a well-collimated large-scale outflow since at least the 1930s \citep[\S\ref{sec:D-bubbleepi};][]{EM98,MR04,MRF07}. Given that protostellar outflows are generally thought to be instrinsically bipolar symmetric, we investigate the possibility that interaction with the ambient medium is obstructing the receding outflow from DG Tau.

Previous studies have considered the interaction of a protostellar outflow with ambient media, which could explain the observed outflow asymmetry. For example, \citet{RC95}, \citet{dGDP99} and \citet{Re02} investigated the effect of a jet impacting a smooth, dense cloud of material, which tended to deflect the jet trajectory. \citet{DFH00} and \citet{WS03} investigated the propagation of spherical winds into a non-spherical protostellar envelope. However, none of these authors replicated the quasi-stationary bubble-like structure we observe in DG Tau.\footnote{Some of the simulations of \citet{dGDP99} do seem to show a small bubble-like structure being formed by the dispersed jet. This may be indicative of the situation occuring at the head of the obscured jet underlying the DG Tau bubble (\S\ref{sec:bubbleAGN-jet}).} In order to find a study which does replicate this structure, we turn to simulations of quasar-mode feedback.

\subsection{Bubbles Driven by AGN Jets}

In hydrodynamic simulations of outflows on the scales of AGN, \cite{SB07} identified four distinct phases of a jet penetrating a two-phase interstellar medium, in which the warm phase material is a clumpy distribution of dense, fractal clouds. Immediately after being launched, the jet enters the \emph{flood-and-channel} phase, as it searches for the path of least resistance through the clouds obstructing it. Then, as the jet head nears the edge of the distribution of clouds, it produces an \emph{energy-driven bubble} that grows larger than the region of obstructing material. As the jet enters the \emph{breakout} phase, it clears the remaining ambient material in its path and nears the edge of the bubble. Finally, the jet pushes through the bubble apex, and enters the \emph{classical} phase, forming the traditional bow-shock morphology at the jet head as it propagates into the wider, more evenly distributed ambient medium \citep[cf.~the large-scale approaching DG Tau outflow;][]{EM98}. \citet{WB11} confirmed this evolution for AGN jets driven through a clumpy ambient medium concentrated near a jet source in an approximately spherical distribution, as is believed to occur in gas-rich protogalaxies. Such a distribution of ambient material is also likely to result from the collapsing protostellar cloud cores that surround early-class protostars. This evolution is well-illustrated in fig.~2 of \citet{WB11}.

Given that the simulations mentioned above deal with AGN jets on scales of kiloparsecs, attempting to apply these results in detail to YSO outflows is inappropriate. However, it is possible to compare the structure of the DG Tau outflows to the evolutionary path described above in a morphological sense. In particular, the morphological similarity that the receding bubble of DG Tau (Figs.~\ref{fig:redchannels}(a), \ref{fig:redvelfit}(a)) bears to the simulations of \citet[][figs.~2 and 3 therein]{SB07}, and \citet[][fig.~2 therein]{WB11}, is remarkable. Furthermore, we note that the velocity field of the warm gas inside the bubble in the simulations of \citet{WB11} (fig.~4 therein) is also consistent with our observations of DG Tau; the highest-velocity material is near the bubble head, with slower material closer to the driving source of the outflow. We also note the velocity of this gas is greater than the expansion velocity of the bubble, as observed in DG Tau (\S\S\ref{sec:obs-fitting}, \ref{sec:analysis-time}). In the case of DG Tau, the gas within the bubble would be stimulated into [Fe II] 1.644 $\mu$m shock-excited line emission by the deceleration of the gas into the bubble walls via shocks, and/or by the turbulence generated by the scattering of the underlying jet.

Before proceeding, we note that the AGN jet-driven bubbles are \emph{energy}-driven, whilst stellar-driven bubbles are typically \emph{momentum}-driven. In both cases, jet material is scattered at the point of obstruction, dissipates to some extent and moves at reduced velocity through the porous medium towards the bubble walls, where it decelerates. In a momentum-driven bubble, the shock-excited material within the bubble walls cools rapidly, so that the bubble expands at a velocity dictated by the conservation of the dispersed jet momentum. However, if the shocked gas does not cool rapidly compared to the dynamical time of the bubble, then the bubble gains a mechanical advantage from the increased thermal pressure and expands more rapidly. This subtle, but important, distinction was first made by \citet{D84}. We propose that the receding outflow of DG Tau is currently in the momentum-driven bubble phase. We present a model consistent with this in \S\ref{sec:model-applic}.

We can determine the evolutionary state of the DG Tau current outflow episode by studying the features of the approaching outflow. Given that this outflow extends beyond the NIFS field, we cannot draw conclusions concerning its evolutionary state based on our data alone. However, \citet{EM98} track the blueshifted outflow out to a bow shock-like structure at $8\farcs 7$ from the central star in the mid- to late-1980s. This is thought to be the extent of the current outflow episode, with previous ejection episodes having driven outflows out to several arcminutes \citep[\S\ref{sec:D-bubbleepi};][]{MR04,MRF07}. This indicates that the approaching outflow has been in the classical phase for at least $\sim 69\textrm{ yr}$ at the 2005 epoch, and is driving into the surrounding ISM. There is no evidence of the remnants of a bubble-like structure in the approaching outflow (\S\ref{sec:model-applic}).

It is generally assumed that collapsing molecular cloud cores are approximately spherically symmetric \citep{SAL87}, and outflow generation in YSOs is often assumed to exhibit bipolar symmetry. In this case, the outflows on either side of the circumstellar disc should evolve at similar rates. This is not observed in DG Tau. We appeal to further AGN outflow simulations to explain this discrepancy. \citet{GKK11} simulated a bipolar AGN jet penetrating a clumpy, gaseous disc on each side of the driving galaxy. Each disc had the same mean density, but a different arrangement of dense clouds within the disc. It was found that this variation in distribution of ambient material results in a difference in time-to-breakout between the two jets of nearly a factor of 4, as a result of clouds being present directly in the jet path on one side of the jet source. We propose such an effect is occurring in DG Tau, and causes an impediment to the propagation of the receding counterjet. This makes it unnecessary to invoke an asymmetry in the jet launching mechanism to explain the differences in morphology between the two outflows.

There is observational evidence for the existence of a clumpy ambient medium above the far-side surface of the DG Tau circumstellar disc. The outermost component of the DG Tau system is a disc-shaped envelope, observed in $^{13}\textrm{CO}$ ($J=1\textrm{--}0$) aperture synthesis data \citep{KKS96a}. This envelope has a radius of $2800\textrm{ AU}$, and a mass of $0.03\textrm{ }M_\odot$, which is calculated from the total $^{13}$CO flux of 40 Jy$\kms$, and by assuming a fractional abundance $X(^{13}\textrm{CO})=10^{-6}$. It is observed to be `clumpy' in nature, and the decrease in $^{13}\textrm{CO}$ column density within $\sim 4000\textrm{ AU}$ of the central star suggests that the outflows from DG Tau have already interacted with the remnant envelope and blown a large portion of it away. We propose that the remnant CO envelope is the clumpy ambient medium impeding the propagation of a receding counterjet. We now search for evidence of the presence of this counterjet.

\subsection{Evidence for a Jet Driving the DG Tau Bubble}\label{sec:bubbleAGN-jet}

A jet is required to provide the energy to drive the bubble expansion as described above. We search for the presence of such a jet in our data, and determine that the morphology of the [Fe II] 1.644 $\mu$m and H$_2$ 1-0 S(1) 2.1218 $\mu$m line emission near the base of the receding outflow structure indicates that a jet-ambient medium interaction is taking place at that location.

We investigate the [Fe II] 1.644 $\mu$m emission-line morphology and kinematics of the receding outflow to search for the presence of a counterjet. The emission knot $\sim 0\farcs 84$ along the outflow axis from the central star does not seem to be a part of the bubble. This knot is located at the same distance to the central star as the tip of the brightest emission region in the approaching jet (Paper I, fig.~3 therein). Furthermore, the fitted line velocity of this knot (170$\kms$, Fig.~\ref{fig:redvelfit}(b)) is equal to the magnitude of the fitted line velocity of the approaching DG Tau jet at a similar distance from the central star (Paper I). Therefore, we interpret this knot to represent a segment of a fast, well-collimated jet emerging from behind the circumstellar disc. The line-velocity map (Fig.~\ref{fig:redvelfit}(b)) shows the presence of a stream of material moving at a line-of-sight velocity $\sim 150\textrm{--}170\kms$ extending from the tip of the fast jet to the apex of the bubble. We interpret this as being indicative of a stream of jet material that is being dispersed by its interaction with the remnant prestellar envelope, and driving the expansion of the bubble.

The time-evolution of the receding outflow from DG Tau (Figs.~\ref{fig:redslices050609}, \ref{fig:redcontours}) may also be interpreted as indicating the presence of an underlying jet. Between the 2005 and 2006 observing epochs, the material that appears to have moved is the highest-velocity emitting material. In the 2009 epoch data, the `apex' of the bubble has disappeared, and a faint trail of high-velocity material can be seen extending from behind the DG Tau disk to the edge of the NIFS field (Fig.~\ref{fig:redslices050609}, [180:220]$\kms$ panel). This may be a signature of the receding DG Tau jet having achieved `breakout' at some point shortly after the 2006 observing epoch. The small movement of the highest-velocity material at the bubble apex corresponds to the outflow beginning to push clear. The base of the bubble remains visible in [Fe II] at the 2009 epoch as a result of the continued cooling of shock-excited material. Further time-monitoring of the DG Tau outflows is required to confirm this suggestion; in particular, the observation of a typical YSO microjet emerging from the bubble remnant would be conclusive evidence.

\begin{figure}
\centering
\includegraphics[width=0.5\textwidth]{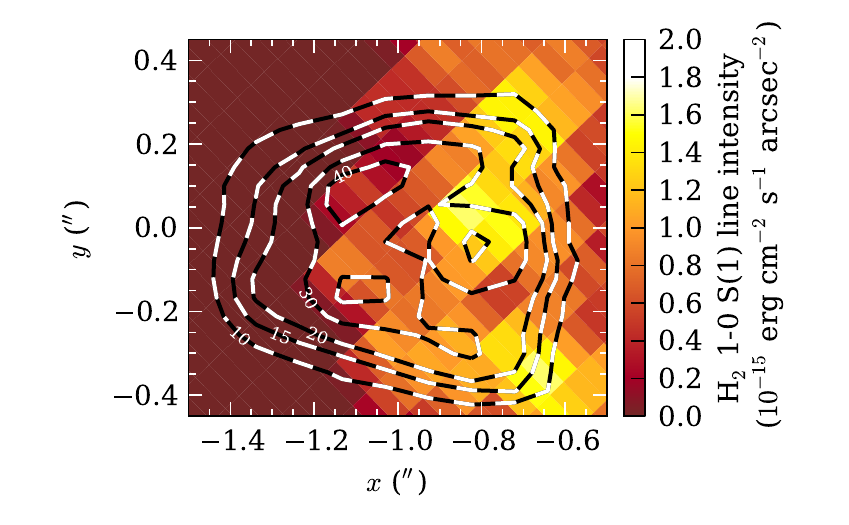}
\caption{H$_2$ 1-0 S(1) 2.1218 $\mu$m integrated emission flux of the receding DG Tau outflow, formed over the velocity range $-100$ to $60\kms$. Overlaid are contours (black with white dashes) of the redshifted [Fe II] 1.644 $\mu$m integrated emission-line flux from the DG Tau outflow (Fig.~\ref{fig:redchannels}(a)), formed over the velocity range $0$ to $340\kms$. Contours are labelled in units of $10^{-15}$ erg cm$^{-2}$ s$^{-1}$ arcsec$^{-2}$.}\label{fig:redH2overlay}
\end{figure}

To search for signs of the jet interacting with the molecular medium above the circumstellar disc surface, we analyse the H$_2$ 1-0 S(1) 2.1218 $\mu$m line emission coincident with the receding outflow \citep{Be08,A-Ae14}. This emission is shown in Fig.~\ref{fig:redH2overlay}, overlaid with contours of the receding DG Tau outflow as seen in [Fe II] 1.644 $\mu$m line emission (Fig.~\ref{fig:redchannels}(a)). The H$_2$ emission is concentrated in three clumps along the base of the bubble. Gaussian line fitting, similar to that performed in \S\ref{sec:obs-fitting}, indicates that the H$_2$ emission is at the systemic velocity. It is unlikely that the material is moving directly across the plane of the sky, so we conclude that this material is stationary with respect to the central star.

That being established, we consider the correlation between the spatial dimensions and positions of the receding DG Tau outflow as seen in [Fe II] 1.644 $\mu$m line emission, and the H$_2$ 2.1218 $\mu$m line emission on the same side of the circumstellar disc. The edges of the bubble structure seen in [Fe II] 1.644 $\mu$m emission are adjacent to two enhancements in H$_2$ emission. The third, central H$_2$ emission enhancement is adjacent to the [Fe II] knot located $0\farcs 84$ from the central star. In view of the systemic velocity of the H$_2$ emission, we propose that it originates from a cloudy ambient molecular medium above the disc surface, which we link to the extended CO envelope around the DG Tau system (\S\ref{sec:bubbleAGN}). This also accounts for the `patchy' nature of the emission compared to the observed blueshifted H$_2$ emission, which emanates from a wide-angle molecular wind \citep[][Paper I]{Be08}. In this picture, the central H$_2$ clump is the cloud that impedes the emergence of the receding DG Tau counterjet, causing the generation of an expanding bubble. The adjacent [Fe II] knot suggests that there is a jet-ambient medium interaction occurring at this point. The H$_2$ clumps at the edge of the bubble correspond to locations where the bubble wall is being driven through the cloud distribution, producing shock-excited emission. \citet{Be08} computed H$_2$ line populations and ratios for the extended emission around DG Tau, and concluded that the emission is stimulated by shock excitation, consistent with our hypothesis. It has previously been suggested that such H$_2$ emission may be stimulated in the wake of a protostellar bow shock \citep{He96}; however, the reader will recall we excluded a bow shock as the cause of the receding outflow structure (\S\ref{sec:bubble}).

Further multi-epoch data are required to confirm this proposed model. If it is correct, then we expect the appearance of a classical YSO microjet from behind the circumstellar disk as the jet continues to push through the inhomogeneous clumpy medium. The intensity of the central H$_2$ 1-0 S(1) 2.1218 $\mu$m emission would diminish as the jet propagates further, although some residual interaction associated with entrainment may continue to excite emission. The bubble emission would then disappear over the course of a cooling time, which we estimate to be $\sim 26\textrm{ yr}$ in \S\ref{sec:model-applic}.

\section{Analytical Modeling}\label{sec:model}

We have proposed that the asymmetry in the DG Tau bipolar outflow is due to the receding counterjet interacting with an inhomogeneous distribution of material above the circumstellar disc surface. This has the effect of dispersing the jet, causing the jet momentum flux to drive the expansion of a bubble structure \citep{SB07,GKK11,WB11}. The evolution of the bubble does not depend on the width or degree of collimation of the original jet; the dispersed secondary flow streams through the porous cloud distribution in all directions \citep{WBU12}. This turbulent, shock-excited dispersed material is the source of the [Fe II] 1.644 $\mu$m line emission observed in DG Tau. We now present an analytical model of the bubble to support our suggestion.

\subsection{Energy-Driven or Momentum-Driven Bubble?}\label{sec:model-applic}

As mentioned above (\S\ref{sec:bubbleAGN}), the bubbles driven by stellar outflows are typically momentum-driven. A characteristic of either an energy- or momentum-driven bubble is a shock front/fronts interior to the bubble walls, where dispersed outflowing material is decelerated to the velocity of the bubble wall \citep{D84,BDO97}. The ratio between the cooling time of this shocked material, and the dynamical time of the bubble, determines if the bubble is energy- or momentum-driven. If the ratio is small (large), then the bubble is momentum-(energy-)driven.

We show that the bubble in the receding DG Tau outflow is momentum-driven by estimating the cooling time of the shocked material inside the bubble, which is defined by the observable redshifted [Fe II] 1.644 $\mu$m line emission. The deprojected velocity observed in the receding outflow $\approx 230\kms$ (\S\ref{sec:obs-fitting}). As this velocity is likely to be much greater than the bubble expansion velocity\footnote{This \emph{a priori} assumption is justified in \S\ref{sec:model-expand}.}, we assume that this is the pre-shock gas velocity\footnote{This model assumes that the outflowing material is stopped in a single strong shock just interior to the bubble walls. It is more likely that this deceleration is gradual, based on the observed velocity profile of the material within the bubble (\S\S\ref{sec:obs-fitting}, \ref{sec:model-expand}). Turbulence may also play a role in stimulating line emission within the bubble. However, this model remains a useful and illustrative approximation.}, $v_1\approx 230\kms$. The post-shock gas temperature, $T_\textrm{sh}$, is then given by
\begin{equation}
T_\textrm{sh} = \frac{3}{16}\frac{\mu m}{k}v_1^2
\end{equation}
\citep{We77}, where $k$ is the Boltzmann constant and $m$ is the atomic mass unit. We take $\mu =1.4$ for a gas having a helium abundance of 10\% of that of hydrogen. The temperature of the post-shock gas, $T_\textrm{sh}= 1.7\times 10^{6}\textrm{ K}$. We further adopt the solar-abundance cooling function, $\Lambda(T)$, of \citet{SD93}, which yields $\Lambda(10^6\textrm{ K})\approx 10^{-22}\textrm{ erg cm}^{3}\textrm{ s}^{-1}$. The electron density within the bubble is $\sim 10^{4}\textrm{ cm}^{-3}$ (\S\ref{sec:obs-density}). We assume that the gas is fully ionised in the hot shocked regions. This allows for the hydrogen number density, $n_\textrm{H}$, to be computed from estimates of the electron density, and the total number density, $n\approx 2.3 n_\textrm{H}$ for total ionisation and the above gas composition. For atomic gas, the gas cooling time, $t_\textrm{c}$, is then given by 
\begin{equation}
t_\textrm{c}=\frac{3}{2}\frac{nkT_\textrm{sh}}{n_\textrm{e}n_\textrm{H}\Lambda(T_\textrm{sh})}\textrm{.}
\end{equation}
Hence, the cooling time of the shocked gas within the bubble is approximately 26 yr.

There are no direct indicators of the age of the bubble. However, if we assume that the DG Tau outflows are driven symmetrically, we may utilise estimates based on the age of the approaching outflow. Observations of the bow shock at the head of the HH 158 complex imply that the current outflow episode in DG Tau commenced circa 1936 \citep{EM98}. This defines the minimum age of the bubble of $69\textrm{ yr}$ at the 2005 epoch. If the blueshifted outflow progressed through the stages of evolution described in \S\ref{sec:bubbleAGN} for a significant period of time, then 1936 represents the approximate breakout date of the blueshifted jet, and the age of the redshifted bubble might then be a factor of a few greater, i.e.~up to a few hundred years. The lack of any visible bubble remnant in the blueshifted outflow suggests that there may not have been a significant momentum-bubble phase on the near side of the circumstellar disc. However, if there was a prolonged bubble phase in the approaching outflow, the lack of a remnant is not surprising, given the short bubble cooling time determined above. Therefore, we are unable to provide a conclusive upper limit to the age of the DG Tau receding bubble. However, by comparing the cooling length determined above to the minimum age of the bubble, we conclude that the cooling time of the bubble is short compared to the bubble dynamical age, and the bubble is therefore momentum-driven.

\subsection{Bubble Evolution}\label{sec:model-eqns}

Let us now consider a momentum-driven bubble model. Let $\dot{M}_\textrm{j}$ and $v_\textrm{j}$ be the mass-loss rate and velocity, respectively, of the jet driving the bubble. The radius of a spherical momentum-driven bubble, $R_\textrm{shell}$, driven by an isotropised outflow into an ambient medium of density $\rho_\textrm{a}$ as a function of bubble age, $t$, is given by
\begin{equation}\label{eq:Rshell}
R_\textrm{shell}(t) = t^{1/2} \left(\frac{\dot{M}_\textrm{j}v_\textrm{j}}{\rho_\textrm{a}}\right)^{1/4} \sqrt{\frac{3}{2\pi}} 
\end{equation}
\citep{D84}. We approximate equation (\ref{eq:Rshell}) to a hemi-ellipsoidal bubble by defining $R_\textrm{shell}(t)$ to be the radius of a spherical bubble having the same volume as the hemi-ellipsoidal bubble.\footnote{\citet{CRA06} provide a full analytical treatment of a non-spherically symmetric wind driving into an ambient medium with a power-law density distribution. However, this detailed model is beyond the needs and scope of this comparison.} This makes $R_\textrm{shell}(t)$ the geometric mean of the bubble height, $x_\textrm{h}(t)$, and the square of the bubble radius, $r_\textrm{c}(t)$, multiplied by a geometric factor, such that
\begin{equation}\label{eq:xhrc}
R_\textrm{shell}(t)=\frac{[x_\textrm{h}(t)r_\textrm{c}^2(t)]^{1/3}}{\sqrt[3]{2}}\textrm{.}
\end{equation}
We introduce a second geometric factor, $f$, describing the elongation of the bubble, such that
\begin{equation}\label{eq:Z}
x_\textrm{h}(t)=fr_\textrm{c}(t)\textrm{.}
\end{equation}
This elongation factor may vary with time. However, on short scales compared to the bubble dynamical age, we assume self-similarity so that it is constant. The height of the bubble as a function of time is then
\begin{equation}\label{eq:xh}
x_\textrm{h}(t)= t^{1/2} \left( \frac{3f^{4/3}}{\pi\sqrt[3]{2}}\sqrt{\frac{\dot{M}_\textrm{j}v_\textrm{j}}{\rho_\textrm{a}}}\right)^{1/2}\textrm{.}
\end{equation}
Then, $x_\textrm{h}\propto f^{2/3}$ and is not very sensitive to the value of $f$.

\subsection{Comparison with Observations}\label{sec:model-compare}

We compare the above model to our observations of the bubble formed by the receding DG Tau outflow. We assume that the outflows driven by DG Tau are symmetric with respect to the circumstellar disc. Therefore, we assign a jet mass-loss rate of $\dot{M}= 5\times 10^{-9}\textrm{ }M_\odot\textrm{ yr}^{-1}$, which is the mass-loss rate of the approaching high-velocity jet (Paper I), to the receding counterjet. We assign a jet velocity, $v_\textrm{j}\approx 230\kms$, based on both the velocity of the inner regions of the approaching jet, as well as the gas velocities observed within the bubble (Fig.~\ref{fig:redvelfit}). 

We determine $f\approx 3.5$ from the observable redshifted emission\footnote{Note that $f\approx 3.5$ is a height-to-radius ratio, unlike the height-to-width ratio $\sim 2$ used in the kinetic models in \S\ref{sec:bubble}.} (Fig.~\ref{fig:redchannels}). The bubble height is measured from the base of the observed structure to the structure apex along the large-scale outflow axis, and is corrected for the inclination of the star-disc system to the line of sight \citep{EM98}. We assume that the bubble major axis is aligned to the large-scale approaching outflow axis. The observed bubble height of $150\textrm{ AU}$ is a minimum value for this parameter. If the bubble extends back to the circumstellar disk surface, it would have a height $\sim 300\textrm{ AU}$. We cannot determine the precise height of the bubble due to obscuration by the circumstellar disc, so we adopt $150\textrm{--}300\textrm{ AU}$ as an acceptable parameter range. The bubble radius is measured perpendicular to the large-scale outflow axis, along the base of the observable structure.

\begin{figure}
\centering
\includegraphics[width=0.5\textwidth]{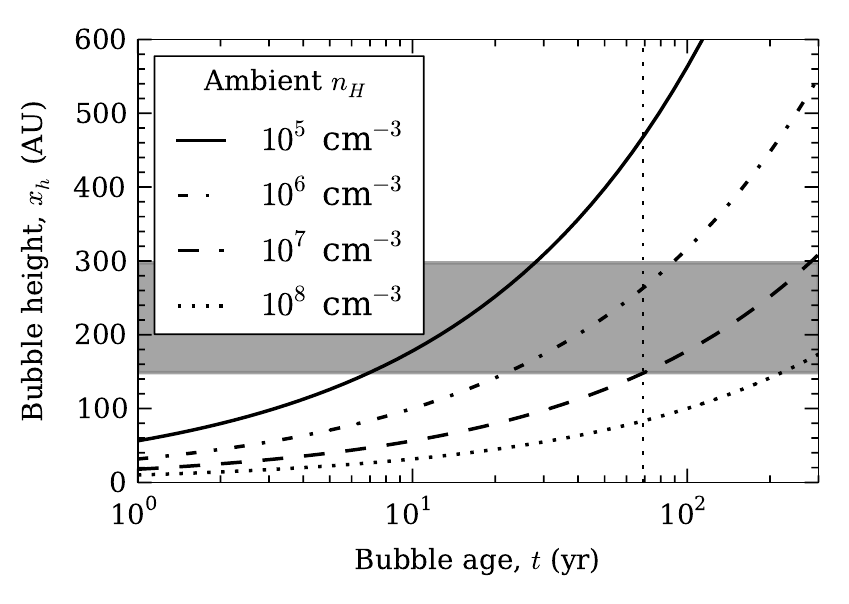}
\caption{Predicted bubble heights, $x_h$, as a function of time for a driving jet of velocity $v\approx 230\kms$ and mass flux $\dot{M}\approx 5\times 10^{-9}\textrm{ }M_\odot\textrm{ yr}^{-1}$, and a bubble of elongation factor $f\approx 3.5$, for various ambient hydrogen number densities (equation (\ref{eq:xh})). A range of possible measured bubble heights, varying between measurement from the base of the observable receding outflow structure, and measurement from the position of the central star, are shown (grey shaded area). The approximate age of the blueshifted outflow as at 2005 is also shown \citep[thin vertical dotted line,][]{EM98}.}\label{fig:bubbleheight}
\end{figure}

The predicted bubble height, $x_\textrm{h}$, as a function of bubble age, $t$, is shown in Fig.~\ref{fig:bubbleheight}. Predicted bubble heights are computed for a range of ambient hydrogen number densities, $n_\textrm{H}$, from $10^5\textrm{ to }10^8\textrm{ cm}^{-3}$. Number densities are converted to mass densities assuming an ambient medium where the helium abundance is 10\% of the hydrogen abundance. These estimates predict that the medium into which the bubble is being driven has a number density of order $n_\textrm{H}\sim 10^6\textrm{--}10^7\textrm{ cm}^{-3}$. This is larger that the average $n_\textrm{H}\sim 10^5\textrm{ cm}^{-3}$ assumed for molecular cloud cores \citep{B11}. However, this higher density is reasonable if the initial cloud core surrounding the star-disc system is still in the process of collapsing onto the circumstellar disc, as is the case with transition Class I/Class II YSOs such as DG Tau. This would increase the density in the central regions of the cloud core, as material concentrates and falls onto the circumstellar disc. 

In DG Tau, the remnant of the cloud core is the CO-emitting envelope surrounding the star-disc system \citep{KKS96a}. We now estimate the density of this envelope. We approximate the envelope shape to be ellipsoidal, with a semi-major axis length of $2800\textrm{ AU}$, and a semi-minor axis length of $160\textrm{ AU}$, corresponding to the height of the observed molecular H$_2$ emission above the central star (\S\ref{sec:bubbleAGN-jet}). Assuming that the envelope is of uniform density, for an envelope mass of $0.03\textrm{ M}_\odot$ \citep[][]{KKS96a}, we calculate its hydrogen number density $n_\textrm{H}\sim 10^{6}\textrm{ cm}^{-3}$ for the above composition. This CO-estimated ambient number density is in agreement with that predicted by our bubble model. Therefore, we conclude that our observations of the DG Tau receding outflow are consistent with a momentum-driven bubble, caused by impediment and dispersion of the receding counterjet by clumpy ambient material in the remnant envelope around the YSO.

\subsection{Bubble Expansion Velocity}\label{sec:model-expand}

\begin{figure}
\centering
\includegraphics[width=0.5\textwidth]{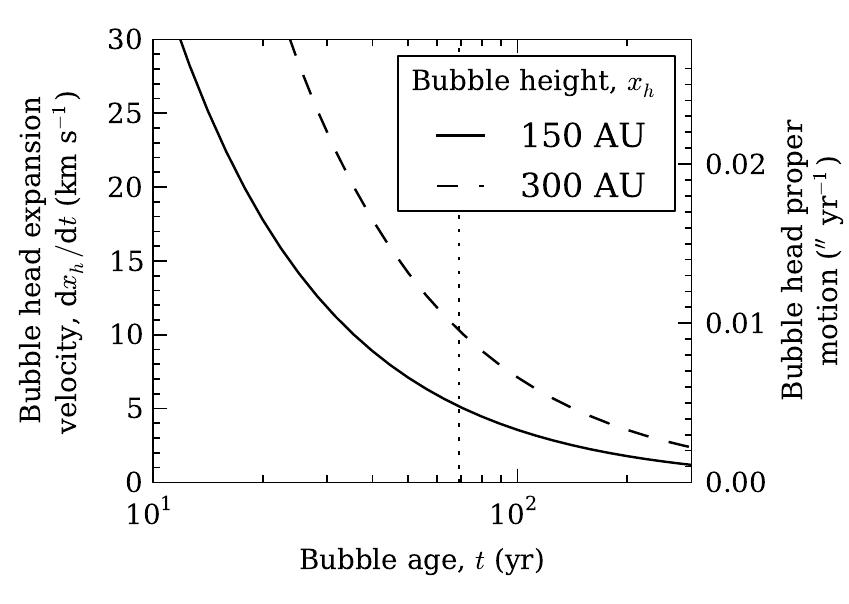}
\caption{Predicted bubble head expansion velocities, $\textrm{d}x_\textrm{h}(t)/\dd t$, as a function of bubble age, $t$, for the two extreme possible bubble heights at the 2005 epoch. The approximate age of the blueshifted outflow at this epoch is denoted by the thin vertical dotted line \citep{EM98}. Bubble head proper motions are calculated assuming that the bubble major axis is aligned to the large-scale outflow axis.}\label{fig:bubbleexpand}
\end{figure}

The bubble model developed above may be used to predict the expansion rate of the bubble. We note that the rate of expansion of the bubble apex can be written as
\begin{equation}\label{eq:exp-velhead}
\frac{\dd x_\textrm{h}(t)}{\dd t} = \frac{1}{2}\frac{x_\textrm{h}(t)}{t}\textrm{.}
\end{equation}
Estimates of the expansion velocity of the bubble head, $\dd x_\textrm{h}(t)/\dd t$, are shown in Fig.~\ref{fig:bubbleexpand}. For the range of possible bubble heights and reasonable bubble ages (\S\ref{sec:model-compare}), the expansion rate of the bubble apex is quite slow, $\lesssim 5\textrm{--}10\kms$. This corresponds to a proper motion on the sky of $\lesssim 0 \farcs 005\textrm{--}0\farcs 01\textrm{ yr}^{-1}$, after accounting for the inclination of the DG Tau disc-jet system to the line of sight \citep{EM98}, and assuming the bubble major axis is aligned to the large-scale outflow axis. As a result of this low expansion velocity, we predict that multi-epoch data covering at least five to ten years are required to accurately measure the bubble expansion rate, and test the accuracy of this model. However, if the jet has already achieved breakout (\S\S\ref{sec:analysis-time}, \ref{sec:bubbleAGN-jet}), this expansion may no longer occur.

\section{Discussion}\label{sec:D}

\subsection{Alternative Mechanisms for the Production of Bipolar Outflow Asymmetry}

Our interpretation of the observed outflow asymmetry in DG Tau as due to the impediment of the receding jet by ambient material does not necessarily exclude other proposed explanations for bipolar outflow asymmetry (\S\ref{sec:intro}). Proposed mechanisms include variations in the near-stellar magnetic field morphology \citep[HD 163296,][]{We06}, a warped circumstellar disc \citep[HH 111, driven by VLA 1,][]{GRL12}, an interaction between a bipolar (quadrupolar) stellar magnetic field and a quadrupolar (bipolar) circumstellar disk magnetic field \citep{Mae12}, or some unspecified mechanism that drives bipolar jets with differing mass-loss rates, but a balance of linear momentum between the two sides of the outflow \citep{Lie12}. We find no evidence through which to exclude any of the proposed mechanisms, predominantly because the obscured counterjet has yet to propagate (and brighten) a sufficient distance from behind the circumstellar disc to measure its physical parameters and kinematics. However, our model of a jet obscured by an asymmetric ambient medium is the only one to explain the \emph{physical} structure of the receding DG Tau outflow. All other models were constructed to explain bipolar \emph{velocity} asymmetry only.

Our model accounts for the velocity asymmetries detected in long-slit spectroscopic data of bipolar YSO outflows \citep[e.g.,][]{Hie94,HMS97}. Based on our data, a long-slit spectrographic observation of DG Tau at the 2005 epoch, with the slit aligned along the outflow axis, would have observed a bipolar outflow velocity asymmetry. At the greatest distance from the central star visible in the NIFS field, the line velocity of the receding outflow is $v_\textrm{r}\sim 180\kms$, whereas the highest-velocity component of the approaching outflow has a line velocity of $v_\textrm{a}\sim 240\kms$ at the same distance from the central star (Paper I). This would appear in a spectroscopic observation of the DG Tau outflows as a velocity asymmetry between the two outflows of $v_\textrm{a}/v_\textrm{r}\sim 1.3$. Although at the lower end of the range of detected velocity asymmetries in bipolar YSO outflows \citep[e.g.,][]{Hie94,HMS97}, it is in agreement with the velocity asymmetry between the DG Tau outflows reported by \citet{Le97}. We suggest spectroimaging follow-up of TTS with bipolar outflow velocity asymmetry would allow structural outflow asymmetry similar to that observed in DG Tau to be detected, and test if the model described above is applicable to other YSOs. We do note that there are some objects, e.g.,~RW Aur, which show outflow velocity asymmetries ($\Delta v\sim 65\kms$ between the two outflows lobes) very close to the central star, and no clear sign of jet/cloud interaction \citep{Woe02}. An alternative model for bipolar velocity asymmetries in such objects is required.

\subsection{Implications for Episodic Ejections}\label{sec:D-bubbleepi}

The approaching HH 158 outflow driven by DG Tau extends $\sim 12\arcsec$ from the central star in 2001--2003 \citep{MRF07}, and terminates in a bow shock that is interpreted to be the head of the current outflow event \citep{EM98}. More recently, \citet{Se03} and \citet{MR04} independently identified another Herbig--Haro complex, HH 702, a $\sim 4\arcmin$-long shock system centred $\sim 10\arcmin$ from DG Tau. This HH complex sits at approximately the same position angle from DG Tau as HH 158, and has been determined to be driven by DG Tau, based on analysis of the proper motions and radial velocities of the knots within the complex \citep{MRF07}. A similar complex, HH 830, was observed at the same distance from DG Tau diametrically opposed to HH 702, but has been determined to not be driven by DG Tau. Hence, the bipolar asymmetry in the DG Tau outflows extends to scales of $\sim 0.5\textrm{ pc}$ from the central star \citep{MR04,MRF07}.

Several other YSOs have also been identified as having associated HH complexes up to $\sim 1\textrm{ pc}$ from the inferred driving source. The location and morphology of these complexes is highly suggestive of periods of relative outflow inactivity, interspersed with mass ejections of various strengths \citep{MRF07}. It has been suggested that this process is linked to the FU Orionis outburst phenomenon, where the optical brightness of a YSO rapidly increases by several orders of magnitude before decreasing back to its original luminosity over a period of $50\textrm{--}100\textrm{ yr}$ \citep{HK96,RA97}. The dynamical time-scales of these parsec-scale YSO outflows were originally thought to match the mean time estimated between FU Orionis outbursts of $10^4\textrm{ yr}$ \citep{H77,HPD03}, although the most recent estimates of these outflow event dynamical time-scales are of order $10^3\textrm{ yr}$ \citep{MRF07}.

It would appear that previous outflow episodes from DG Tau have failed to clear a path through the ambient medium on the far side of the circumstellar disc, given the lack of associated Herbig--Haro objects at large distances on the receding side of the outflow, and the nature of the current receding outflow close to the circumstellar disc. The shape of the extended CO envelope suggests that a significant amount of ambient material has been driven off by the approaching outflow, but not the receding outflow, which suggests the receding outflow has failed to clear the envelope. However, it is possible that a large-scale redshifted outflow does exist, but is hidden from view as it recedes into the large-scale molecular cloud behind DG Tau \citep{MRF07}. Such an outflow would have cleared a channel through the ambient material immediately above the disc in order to escape. Indeed, the $^{13}$CO(2-1) map of \citet{Tes02} seems to indicate a channel may have been cleared through the material around DG Tau by a receding outflow. If so, this raises the question of why the present receding outflow is not using the same cleared channel, and propagating to the same distance from the central star as the approaching outflow.

We determine that an outflow channel formed through the extended envelope immediately above the circumstellar disc \citep{KKS96a} could close over between ejection events, based on simple estimates. The dynamical age of the observed large-scale DG Tau outflow is estimated to be $\sim 2.3\times 10^3\textrm{ yr}$ \citep{MRF07}, which leads to a time between high-mass flux ejection events of $\sim 2.2\times 10^3\textrm{ yr}$, assuming the most recent ejection event began in 1936 \citep{EM98}. Based on the length of the HH 702 complex of $4\arcmin$, and the average proper motion of the HH 702 knots of $\sim 175\kms$, we estimate the dynamical time of the previous high-mass flux outflow event to be $\sim 900\textrm{ yr}$. This leaves a time period of $\sim 1.3\times 10^3\textrm{ yr}$ during which the outflows from DG Tau would be comparatively weak. 

Let us suppose that the receding outflow cleared a path through the extended envelope above the circumstellar disc during a previous ejection event. We assume that the width of this channel is $\sim 40\textrm{ AU}$, corresponding to the largest jet diameter observed in the approaching DG Tau microjet (Paper I). We estimate the sound speed in the molecular cloud core to be $\sim 0.3\kms$, assuming a typical molecular cloud core temperature of $\sim 10\textrm{ K}$ \citep{B11} and a molecular medium with adiabatic index $\gamma= 7/5$. The crossing time over the width of the channel is then $\sim 630\textrm{ yr}$. The period of inactivity in the outflows is $\gtrsim 2$ times the crossing time of the channel, indicating the channel may become closed over during the period of relative quiescence. Therefore, we conclude that either side of the DG Tau outflow may be required to go through a bubble-driving phase for \emph{any} ejection event. Whether this occurs or not is completely dependent on whether the motions of the ambient material above the circumstellar disc close the channel created by the previous outflow event. As mentioned above (\S\ref{sec:model-applic}), the cooling time in these bubbles is short, so there would` be little observable remnant of the bubble within a few decades after the jet achieves break-out.

\section{Conclusions}\label{sec:concl}

We have investigated the nature of the receding microjet-scale outflow of the YSO DG Tauri, utilising high-resolution spectroimaging data taken with the Near-infrared Integral Field Spectrograph (NIFS) on Gemini North. In the $H$-band, the outflow appears as a large bubble-like structure in [Fe II] 1.644 $\mu$m line emission. This is in stark contrast to the approaching outflow, which shows a two-component outflow, consisting of a high-velocity jet surrounded by a lower-velocity disc wind, stimulated into emission by turbulent entrainment (Paper I). In the $K$-band, `clumpy' H$_2$ 1-0 S(1) 2.1218 $\mu$m line emission is observed near the edge of the circumstellar disc, coincident with some of the brightest regions of redshifted [Fe II] emission. Line fits of this emission showed that the H$_2$-emitting gas is at the systemic velocity, indicating that it represents material stationary with respect to the central star instead of an outflow component.

The emission-line velocity structure of the receding outflow is well-described by kinetic models of bubbles with an internal distribution of expanding, radiating gas. These models generate simulated emission line channel maps from input parameters such as bubble height, elongation, distance to the outflow source, and inclination of the outflow axis to the line of sight. Simulated IFU data produced by the models show our observational data are consistent with the presence of a stationary bubble, with an internal distribution of emitting gas expanding towards the bubble walls.

We compared the current appearance of the receding DG Tau bubble with the four-stage evolutionary track for AGN jet-driven bubbles proposed by \citet{SB07}, and further refined by \citet{WB11}. We concluded that the DG Tau receding outflow is forming a momentum-driven bubble at the 2005 observing epoch, morphologically similar to the energy-driven bubble phase identified by \citet{SB07}. The receding counterjet is being blocked by a cloud of ambient medium in its path. As the jet begins to push past the clump, it blows a large bubble, which extends further than the distribution of material that is blocking its progress. We interpreted the observed H$_2$ emission as being indicative of such a clumpy medium above the surface of the circumstellar disc, which is interacting with both the jet and the expanding bubble. This interpretation is supported by the presence of an [Fe II] emission enhancement adjacent to one of the H$_2$ clumps, which suggests a jet-ambient medium interaction point. The presence of a clumpy, large-scale residual envelope of molecular material around DG Tau has been observed previously \citep{KKS96a}, lending further weight to this interpretation. 

Our multi-epoch data support this interpretation. Between 2005 and 2006, the bubble is effectively stationary, except for a small ($\sim 70\textrm{--}90\kms$) movement of the highest-velocity emitting material at the bubble apex. In 2009, this high-velocity material has disappeared. This may be indicative of the jet driving the bubble achieving breakout at some point after the 2006 observing epoch, leaving just the base of the bubble observable in the 2009 epoch. 

We constructed an analytical model of a jet momentum-driven bubble. This model describes the evolution of a bubble driven by a dispersed jet, propagating into a smooth ambient medium. Based on the physical scale of the receding DG Tau bubble, this model predicts that the ambient medium number density above the circumstellar disc surface is of order $10^6\textrm{ to }10^7\textrm{ cm}^{-3}$, which is in agreement with density estimates of the extended CO envelope around DG Tau. The model predicts a bubble expansion of $\lesssim 5\textrm{--}10\kms$ at the bubble head, which can be tested with multi-epoch observational data covering a five- to ten-year period. The cooling time of the bubble is estimated to be $\sim 26\textrm { yr}$. This is the time-scale on which the bubble disappears due to cooling once the jet head moves beyond the apex of the bubble. This also implies that the bubble is momentum-driven, as this cooling time is short compared to the estimated dynamical age of the bubble.

Our conclusions do not necessarily exclude other proposed models for YSO bipolar outflow asymmetry. However, an asymmetric ambient medium obstructing the evolution of a symmetric bipolar outflow is the only model that explains the morphology of the receding DG Tau outflow. We have demonstrated that a long-slit spectrographic observation of DG Tau would weakly replicate the bipolar outflow velocity asymmetry observed in long-slit spectrographic observations of other YSOs, and is in agreement with the previous spectroscopic measurement for DG Tau \citep{Le97}. We have also shown that the outflow channel formed by an episode of outflow activity may close over during periods of relative outflow quiescence, based on observations of the large-scale DG Tau approaching outflow $\lesssim 0.5\textrm{ pc}$ from the central star. This implies that both the approaching and receding outflows from DG Tau may be forced to evolve through a bubble phase for any given outflow event.

Multi-epoch data of DG Tau is the best method to test these conclusions. Our analysis has provided robust predictions of the evolutionary path that the bubble will take once the underlying jet breaks free of the ambient medium that obstructs its progress. These predictions may be directly compared to future observations.

\section*{Acknowledgements}

Based on observations obtained at the Gemini Observatory, which is operated by the 
Association of Universities for Research in Astronomy, Inc., under a cooperative agreement 
with the NSF on behalf of the Gemini partnership: the National Science Foundation 
(United States), the National Research Council (Canada), CONICYT (Chile), the Australian 
Research Council (Australia), Minist\'{e}rio da Ci\^{e}ncia, Tecnologia e Inova\c{c}\~{a}o 
(Brazil) and Ministerio de Ciencia, Tecnolog\'{i}a e Innovaci\'{o}n Productiva (Argentina).

We are extremely grateful for the support of the NIFS teams at
the Australian National University, Auspace, and Gemini Observatory for their tireless efforts during the instrument integration, commissioning and system verification: Jan Van Harmleen, Peter Young, Mark Jarnyk (deceased), Nick Porecki, Richard Gronke, Robert Boss, Brian Walls, Gelys Trancho, Inseok Song, Chris Carter, Peter Groskowski, Tatiana Paz, John White, and James Patao.

We thank the anonymous referee for his/her comments on the manuscript. We wish to thank A.~Y.~Wagner for fruitful discussions on the nature of jet-driven bubbles, and for useful comments on a draft version of this paper. M.~White acknowledges the generous travel support from Academia Sinica to attend the conference Star Formation Through Spectroimaging at High Angular Resolution in July 2011, which provided useful information for this study. This work was supported by the Australian Research Council through Discovery Project Grant DP120101792 (R.~Salmeron).

\bibliographystyle{mn2e}
\bibliography{library}

\end{document}